\documentclass[showpacs, preprint, superscriptaddress, amsmath, amssymb, aps, prmaterials]{revtex4-2}
\usepackage{graphicx}% Include figure files
\usepackage{dcolumn}% Align table columns on decimal point
\usepackage{bm}% bold math
\usepackage[utf8]{inputenc}

\usepackage[colorlinks=true, linkcolor=blue, urlcolor=blue, citecolor=blue]{hyperref}

\usepackage{color}
\usepackage{xcolor}
\usepackage{siunitx}
\DeclareSIUnit{\rpm}{rpm} % Defines 'rpm' as a unit
\DeclareSIUnit{\torr}{Torr} % Define the Torr unit
\usepackage[version=4]{mhchem}
\usepackage{gensymb}
\DeclareRobustCommand{\rchi}{{\mathpalette\irchi\relax}}
\newcommand{\irchi}[2]{\raisebox{\depth}{$#1\chi$}}
\emergencystretch=\maxdimen
\begin{document}

\title{Single-crystalline CrSb(0001) thin films grown by dc magnetron co-sputtering}

\author{S. P. Bommanaboyena}
\affiliation{Institute of Physics, Czech Academy of Sciences, Cukrovarnická 10, 162 00 Prague 6, Czech Republic}
\author{C.~Müller}
\affiliation{Institute of Physics, Czech Academy of Sciences, Cukrovarnická 10, 162 00 Prague 6, Czech Republic}
\affiliation{Physics Department, University of Regensburg, 93040 Regensburg, Germany}
\author{M.~Jarošová}
\affiliation{Institute of Physics, Czech Academy of Sciences, Cukrovarnická 10, 162 00 Prague 6, Czech Republic}
\author{K.~Wolk}
\affiliation{Institute of Physics, Czech Academy of Sciences, Cukrovarnická 10, 162 00 Prague 6, Czech Republic}
\affiliation{Department of Materials Science and Engineering, Norwegian University of Science and Technology (NTNU), 7034 Trondheim, Norway}
\author{S.~Telkamp}
\affiliation{Solid State Physics Laboratory, ETH Zurich, 8093 Zurich, Switzerland}
\author{P.~Zeng}
\affiliation{Scientific Centre for Optical and Electron Microscopy (ScopeM), 8093 ETH Zurich, Switzerland}
\author{F.~Křížek}
\affiliation{Institute of Physics, Czech Academy of Sciences, Cukrovarnická 10, 162 00 Prague 6, Czech Republic}
\affiliation{Center for Quantum Devices, Niels Bohr Institute,
University of Copenhagen, 2100 Copenhagen, Denmark}
\author{\mbox{T. Uchimura}}
\affiliation{Laboratory for Nanoelectronics and Spintronics, Research Institute of Electrical Communication, Tohoku University, Sendai 980-8577, Japan}
\affiliation{Graduate School of Engineering, Tohoku University, Sendai 980-0845, Japan}
\author{A.~Badura}
\affiliation{Institute of Physics, Czech Academy of Sciences, Cukrovarnická 10, 162 00 Prague 6, Czech Republic}
\affiliation{Department of Condensed Matter Physics, Charles University,\\Ke Karlovu 5, 121 16 Prague, Czech Republic}
\author{K.~Olejník}
\affiliation{Institute of Physics, Czech Academy of Sciences, Cukrovarnická 10, 162 00 Prague 6, Czech Republic}
\author{D.~Scheffler}
\affiliation{Institute of Physics, Czech Academy of Sciences, Cukrovarnická 10, 162 00 Prague 6, Czech Republic}
\author{K.~Beranová}
\affiliation{Institute of Physics, Czech Academy of Sciences, Cukrovarnická 10, 162 00 Prague 6, Czech Republic}
\author{S.~Banerjee}
\affiliation{Institute of Physics, Czech Academy of Sciences, Cukrovarnická 10, 162 00 Prague 6, Czech Republic}
\affiliation{Faculty of Nuclear Sciences and Physical Engineering, Czech Technical University in Prague, Trojanova 13, 120 00 Prague 2, Czech Republic}
\author{M.~Ledinský}
\affiliation{Institute of Physics, Czech Academy of Sciences, Cukrovarnická 10, 162 00 Prague 6, Czech Republic}
\author{H.~Reichlová}
\affiliation{Institute of Physics, Czech Academy of Sciences, Cukrovarnická 10, 162 00 Prague 6, Czech Republic}
\author{T.~Jungwirth}
\affiliation{Institute of Physics, Czech Academy of Sciences, Cukrovarnická 10, 162 00 Prague 6, Czech Republic}
\affiliation{School of Physics and Astronomy, University of Nottingham, Nottingham NG7 2RD, United Kingdom}
\author{L.~Horák}
\affiliation{Department of Condensed Matter Physics, Charles University,\\Ke Karlovu 5, 121 16 Prague, Czech Republic}
\author{D.~Kriegner}
\email{kriegner@fzu.cz}
\affiliation{Institute of Physics, Czech Academy of Sciences, Cukrovarnická 10, 162 00 Prague 6, Czech Republic}

\date{\today}

\begin{abstract}
The recent discovery of altermagnetism has sparked renewed interest in the growth of epitaxial films of the NiAs-phase polymorph of CrSb. This paper describes the magnetron sputtering-based fabrication and characterization of high-quality single crystalline CrSb(0001) thin films supported by an isostructural non-magnetic PtSb buffer. X-ray diffraction and scanning transmission electron microscopy show that the films are phase-pure and possess a very high crystalline quality (mosaicity $\approx0.05\degree$), while also being free of extended crystallographic defects. Both scanning electron microscopy and atomic force microscopy confirm their smooth and homogeneous topography. Additionally, the elemental composition of our films was found to be close to stoichiometric via electron probe microanalysis and X-ray fluorescence. Thus, the developed samples represent an ideal platform for further investigation of the material properties of CrSb.
\end{abstract}

\maketitle

\section{Introduction}

Over the past decade, compounds with a fully compensated magnetic structure have gained significant attention for spintronics applications due to their advantageous properties, including ultrafast magnetization dynamics, stability in external magnetic fields and the absence of stray fields \cite{Jungwirth2016, Baltz2018}. This interest has been further intensified by the recent theoretical prediction \cite{Smejkal2022, Smejkal2022dec} and subsequent experimental confirmation of altermagnetism \cite{Krempaský2024, Ding2024, Lee2024, Osumi2024, Reimers2024, li2024, Zeng2024, Yang2025, Amin2024}. 
Altermagnetic materials exhibit collinear-compensated magnetic ordering, characterized in the limit of zero spin-orbit coupling by (some) spontaneously broken continuous spin-space rotation symmetries and (some) real-space rotation symmetries of the crystallographic point group \cite{Smejkal2022, Smejkal2022dec}. Simultaneously, these materials retain a symmetry combining crystallographic real-space rotation with time reversal \cite{Smejkal2022Mar, Smejkal2022, Smejkal2022dec}. The anisotropic spin-polarized band structures of 3D altermagnets are classified as $d$-wave, $g$-wave or $i$-wave with respect to 2, 4 or 6 nodal planes crossing the $\Gamma$-point in the Brillouin zone. These nodal planes feature symmetry-protected spin degeneracy in the limit of zero spin-orbit coupling \cite{Smejkal2022Mar, Smejkal2022, Smejkal2022dec}. Consequently, altermagnets host a variety of effects previously thought to be forbidden in collinear compensated magnets. These include the anomalous Hall effect (AHE) \cite{Smejkal2020, Smejkal2022Mar, Betancourt2023, Reichlova2024, Leiviska2024, Rial2024} as well as X-ray magnetic circular dichroism (XMCD) \cite{Hariki2024}, which have attracted considerable interest. For example, XMCD, in combination with linear dichroism and photoemission, enabled a high-resolution direct-space vector mapping of Néel vector domains and nano-textures in the altermagnet MnTe \cite{Amin2024}. The anisotropic nature of altermagnetism and its strong connection to symmetry suggest that high-quality crystals are best suited for its investigation. Owing to its large spin splitting \cite{Smejkal2022, Smejkal2022dec}, CrSb has garnered widespread recognition as a promising altermagnet \cite{Reimers2024, Ding2024, Urata2024, li2024, Zhou2025, Zeng2024, Yang2025}. CrSb is isostructural to altermagnetic MnTe \cite{Krempaský2024, Betancourt2023, Hariki2024, Amin2024}, but unlike MnTe, it is metallic with a strong uniaxial magnetic anisotropy \cite{Snow1952, Reimers1982, Yuan2020} and high Néel temperature of $\approx$\SI{700}{\kelvin} \cite{Willis1953, LOTGERING1957, Reimers1982}.

Both altermagnetic CrSb and MnTe are hexagonal with the crystallographic space group $P6_3/mmc$ and spin point group $^26 / ^2m ^2m ^1 m$. They belong to the NiAs structure type (B8$_1$), similar to MnSb \cite{TERAMOTO1968, Okita1968, Coehoorn1980}, a ferromagnet well explored in thin film form. For epitaxial growth of NiAs compounds in the (0001) \footnote{Note that we use Miller indices $hkl$ to denote cubic lattice points/directions and Bravais-Miller indices $hkil$ ($i = -h - k$) to denote hexagonal lattice points/directions.} ($c$-plane) orientation, the (111) plane of zinc-blende semiconductors has been the traditional substrate of choice due to its 6-fold rotational symmetry \cite{TATSUOKA1996, HATFIELD2007, ALDOUS2012, Kriegner2016, Amin2024}. However, native oxide removal from III-V semiconductor surfaces is challenging and best performed in situ via thermal desorption under a controlled \mbox{group-V} flux to compensate for inevitable losses. Akinaga et al. even found that a freshly deposited homoepitaxial buffer on GaAs(111)B was crucial for MnSb(0001) film growth \cite{Akinaga1997}. Unfortunately, such complex surface preparation often limits the high-quality growth of NiAs-type crystals to molecular beam epitaxy (MBE) systems equipped for this purpose.

In this report, we show the growth of single-crystalline (0001)-oriented CrSb films on a lattice-matched \ce{SrF2}(111)/PtSb(0001) template via direct current (dc) magnetron co-sputtering. \ce{SrF2} is a comparatively inert insulator with the \ce{CaF2} crystal structure, allowing for much simpler surface preparation. PtSb, our NiAs-structured buffer material, is a non-magnetic, superconducting metal ($T_c\sim\SI{2.1}{\kelvin}$) \cite{Matthias1953} that we find to grow epitaxially with an atomically flat surface on \ce{SrF2}. Using this buffer layer, we seed the growth of CrSb in the same crystallographic orientation. In this paper, we provide a comprehensive analysis of the chemical phase, crystallographic orientation, mosaicity and morphology of our heterostructure. We also demonstrate the efficacy of our method over previous growth recipes, discuss the crucial role of the buffer layer and present the basic magnetotransport properties of our thin films.

\section{Fabrication of P\lowercase{t}S\lowercase{b}/C\lowercase{r}S\lowercase{b} bilayers}

\begin{figure}
    \centering
    \includegraphics[width=0.70\textwidth, keepaspectratio]{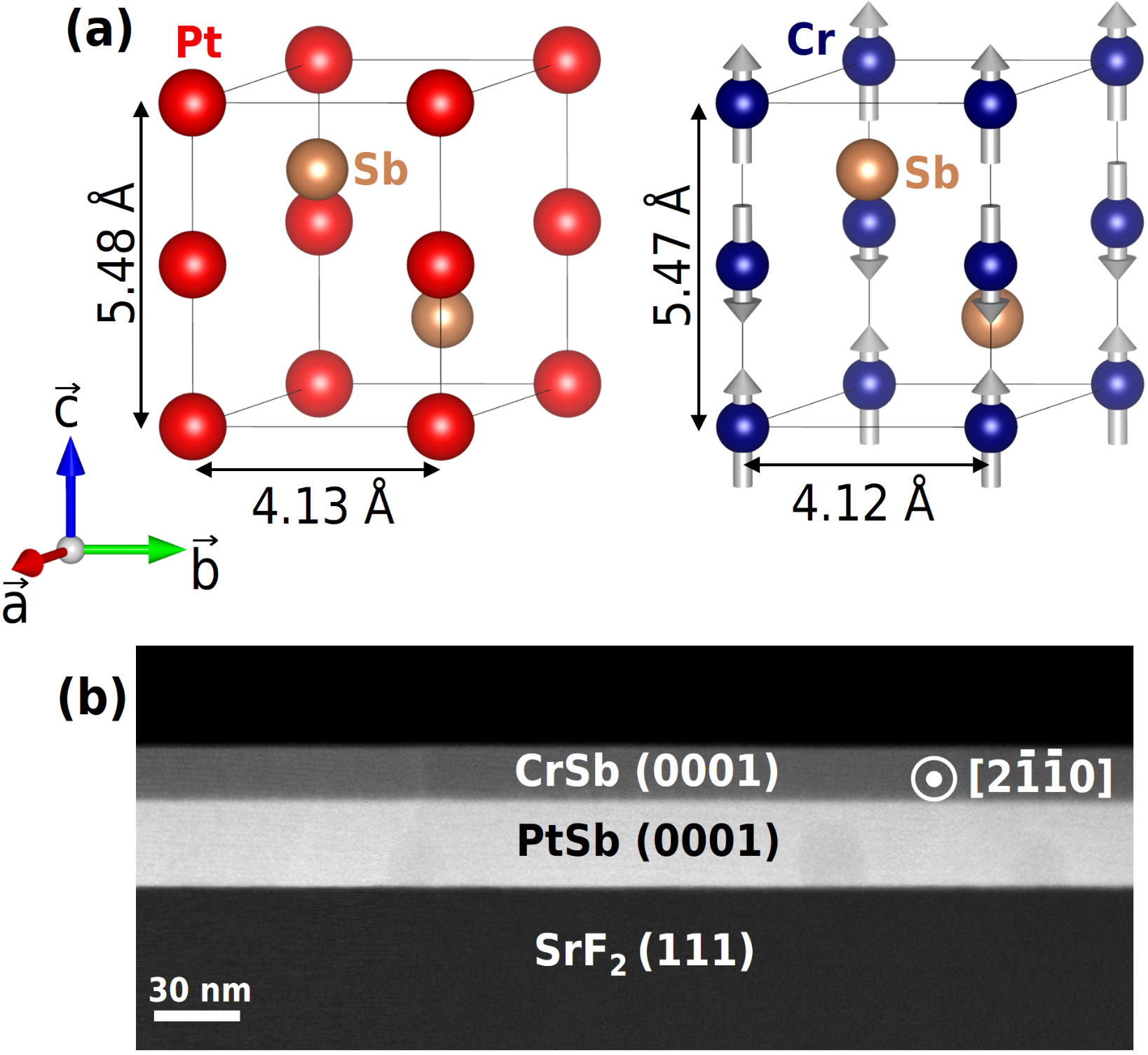}
    \caption{(a) Unit cells of PtSb and CrSb with lattice parameters in bulk crystals. Arrows attached to Cr atoms indicate the magnetic order. (b) Cross-sectional HAADF-STEM image of a polished \ce{SrF2}(111) (substrate)/\SI{30}{\nm} PtSb(0001) (buffer)/\SI{20}{\nm} CrSb(0001) heterostructure with zone axis $[2\bar1\bar10]$.}
    \label{Unit cells}
    \label{xTEM}
\end{figure}

The thin film growth was performed in a confocal magnetron sputtering system with four sputtering guns designed to accommodate 2-inch targets. The deposition chamber has a base pressure $<$~\SI{3e-8}{\milli\bar} and the working gas (99.999\% pure Ar) pressure used for the sputtering of both layers was \SI{6.5e-3}{\milli\bar}. 99.99\% pure elemental targets of Pt, Cr and Sb were used for this study and their individual sputtering fluxes were measured by an in-situ oscillating quartz crystal monitor. Pt:Sb and Cr:Sb flux ratios were adjusted to achieve optimal stoichiometry in each layer (explained further in the next section). \mbox{$10\times10\times\SI{0.5}{\milli\meter^3}$} \ce{SrF2}(111) substrates with edges parallel to $[1\bar10]$ and $[11\bar2]$ were supplied by CrysTec GmbH. Both polished and freshly cleaved substrates were used. CrSb and PtSb are hexagonal (Fig.~\hyperref[Unit cells]{\ref*{Unit cells}(a)} \cite{Momma2008}), with lattice constants in the vicinity of $a = $ \SI{4.12}{\angstrom}, $c = $ \SI{5.47}{\angstrom} \cite{Reimers1982, LOTGERING1957, Kjekshus1969, Kallel1974} and $a =$ \SI{4.13}{\angstrom}, $c =$ \SI{5.48}{\angstrom} \cite{Kjekshus1969, Zhuravlev1962, Ellner2004}, respectively. Therefore, the (111) surface of cubic \ce{SrF2} ($a =$ \SI{5.80}{\angstrom}), being composed of hexagons with a side of $a/\sqrt{2} =$ \SI{4.10}{\angstrom}, serves as an well-matched substrate with a small in-plane lattice mismatch of $\sim0.5\%$. The substrate was first thoroughly cleansed using isopropanol at \SI{60}{\celsius} in an ultrasonic bath for 5 minutes. Subsequently, it was annealed at \SI{650}{\celsius} \footnote{Note that the heater temperatures given here were measured by an in-situ thermocouple.} in the growth chamber for 45 minutes prior to deposition to remove adsorbed water and remaining organic contamination from the surface. The temperature was then set to \SI{400}{\celsius}, after which Pt and Sb were co-sputtered onto the hot substrate to grow a PtSb(0001) seed crystal. The sample holder was maintained at a height of $\sim$~\SI{170}{\milli\meter} from the targets and was rotated at a constant speed of \SI{40}{\rpm} throughout the deposition process to ensure homogeneous layers. Further, the sample temperature was lowered to \SI{200}{\celsius} for the co-deposition of Cr and Sb. The temperature values mentioned here were optimized to achieve PtSb(0001) and CrSb(0001) with the best crystal quality while avoiding interdiffusion of the constituent elements across these two layers. Using this process, smooth and distinct films can be grown on the substrate, as evidenced by high-angle annular dark-field scanning transmission electron microscopy (HAADF-STEM) of a cross-sectional lamella (Fig.~\hyperref[xTEM]{\ref*{xTEM}(b)}). Additionally, samples with CrSb deposited directly on InP(111) A and B surfaces ($a/\sqrt{2} =$ \SI{4.15}{\angstrom}), GaAs(111)B ($a/\sqrt{2} =$ \SI{4.00}{\angstrom}) and \ce{SrF2}(111) at various temperatures were prepared for comparison.

\section{Characterization}

\begin{figure*}
    \centering
    \includegraphics[width=\textwidth, keepaspectratio]{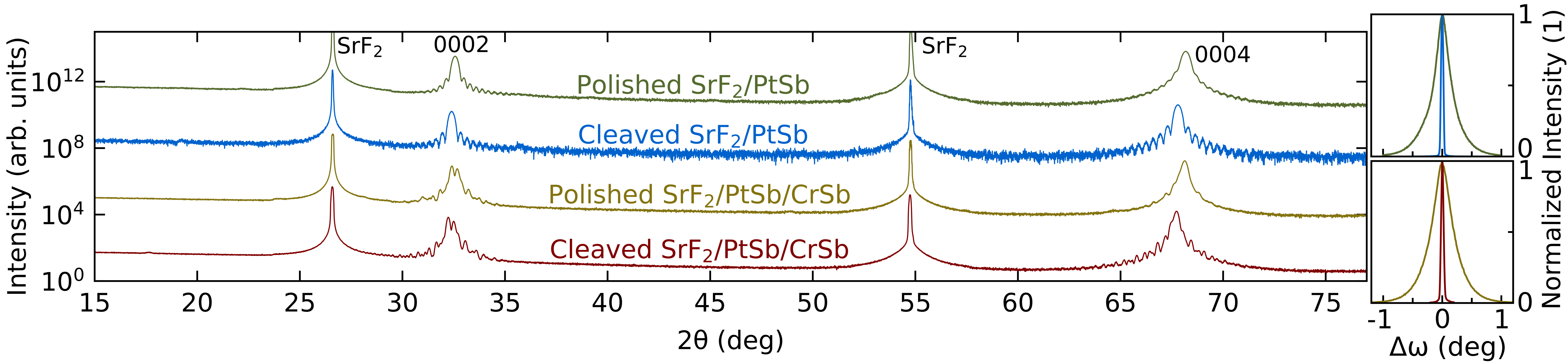}
    \caption{X-ray diffraction symmetric radial scans of our optimized films, performed in the parallel beam geometry with the detector in scanning 1D mode and using Cu-K$\alpha_1$ radiation. The results are presented for films grown on both cleaved and polished \ce{SrF2}(111) substrates and are vertically offset for clarity. The corresponding rocking curves of the 0004 Bragg peak, measured with the detector in 0D mode, are presented on the right side (using the same line colors). The curves show a full width at half maximum (FWHM) of $0.05\degree$ and $0.35\degree$ in cleaved and polished cases, respectively.}
    \label{radial}
\end{figure*}

\begin{figure}
    \centering
    \includegraphics[width=0.8\textwidth, keepaspectratio]{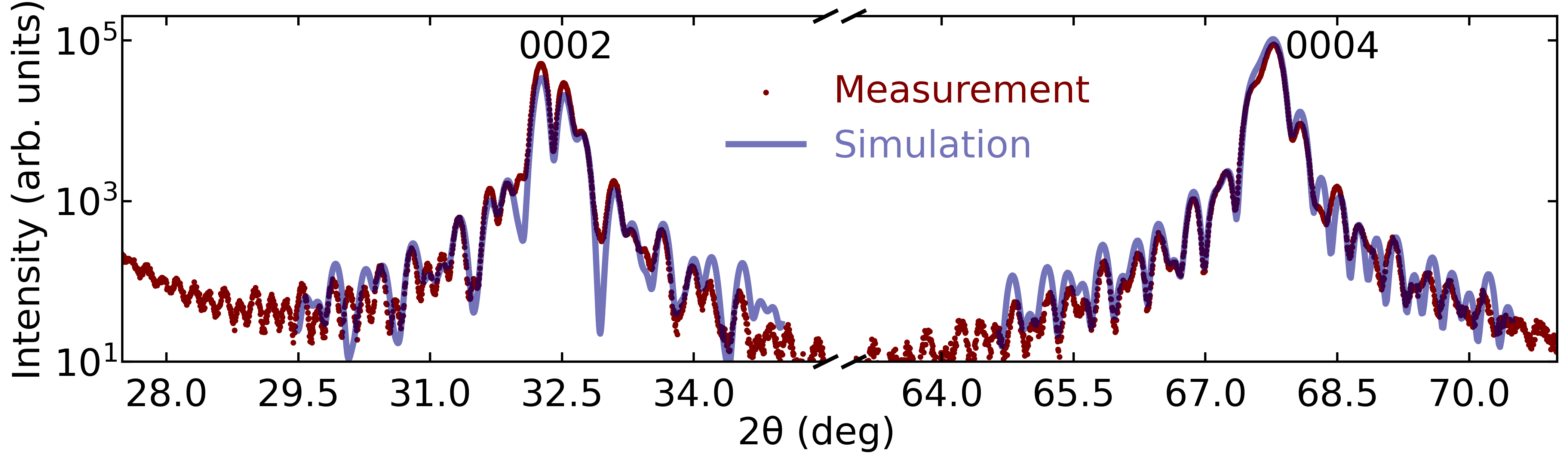}
    \caption{0002 and 0004 film peaks from the XRD symmetric radial scan of cleaved \ce{SrF2}(111)/\SI{30}{\nm} PtSb(0001)/\SI{20}{\nm} CrSb(0001), with the detector in 0D mode. The diffraction signals can be simulated well using the kinematic multibeam model of diffraction.}
    \label{Simulations}
\end{figure}

X-ray diffraction (XRD) symmetric radial scans were carried out on our thin films to determine their crystallographic structure and orientation. Despite the favorable lattice match, CrSb grows in a polycrystalline manner when deposited directly on all the substrates mentioned in the previous section (see Fig.~S1 of Supplemental Material \cite{Supplemental_material}). (0001)-oriented crystallites are partly favored on \mbox{III-Vs} but absolutely no sign of these is found on \ce{SrF2}(111). Changing the deposition temperature was found to have little to no influence on the crystallite orientations that typically form. Contrary to CrSb, PtSb is observed to be single crystalline on the same substrate with its [0001] axis perpendicular to the surface plane as expected (Fig.~\ref{radial}). Both 0002 and 0004 peaks show multiple orders of Laue oscillations, implying a very high crystalline quality. This is further corroborated by the rocking curve widths of the PtSb 0002 and 0004 Bragg peaks, whose combined analysis enables the extraction of the mosaicity, yielding values of $\approx0.05\degree$ and $0.35\degree$ on cleaved and polished \ce{SrF2}, respectively (substrate mosaicity = $0.01\degree$ – $0.02\degree$ in both cases). Using higher deposition temperatures was found to cause pinholes in the PtSb layer. The diffraction peaks corresponding to the PtSb/CrSb bilayer grown on a cleaved substrate are rather complex due to the overlap of diffraction signals from the two layers. The signals show several orders of prominent Laue oscillations for both diffraction orders. They can be simulated with the kinematic multi-beam model of diffraction implemented in \emph{xrayutilities} \cite{Kriegner2013} and is presented in Fig.~\ref{Simulations}. The observed peaks correspond to the 0002 and 0004 Bragg peaks of the combined PtSb/CrSb system. This close agreement of simulation and measurement allows us to conclude that the CrSb epilayer is also single crystalline and (0001)-oriented as desired. The rocking curve of the combined 0004 Bragg peak has a FWHM of 0.05$\degree$. Since the PtSb film showed the same width, we infer that the CrSb layer  inherits the low mosaicity of the seed layer, exhibiting similar crystal quality. Based on the fit in Fig.~\ref{Simulations}, $c$-axis lattice constants of \SI{5.53}{\angstrom} and \SI{5.51}{\angstrom} are extracted for PtSb and CrSb, respectively. A reciprocal space map of the vicinity of the PtSb/CrSb~$11\bar24$ Bragg spot allows us to determine $a =$ \SI{4.10}{\angstrom}, thereby implying an in-plane compressive strain of -0.5\% (all reciprocal space maps are provided in Fig.~S2 of Supplemental Material \cite{Supplemental_material}). This indicates that the film adopts the in-plane lattice spacing of the \ce{SrF2} substrate, suggesting growth with no misfit dislocations. The same bilayer was found to grow with fewer and less pronounced Laue oscillations and a mosaicity of 0.35$\degree$ on polished \ce{SrF2}(111) (Fig.~\ref{radial}). Here, $c =\SI{5.50}{\angstrom}$ and \SI{5.49}{\angstrom} were extracted for PtSb and CrSb, respectively. Remarkably, unlike the previous case, these layers are relaxed in-plane with $a =$~\SI{4.12}{\angstrom}. This is consistent with their smaller $c$ values and aligns well with the reported bulk lattice parameters. The relaxation is likely the result of enhanced dislocation nucleation at the roughness present on polished substrate surfaces.

\begin{figure}
    \centering
    \includegraphics[width=0.8\textwidth, keepaspectratio]{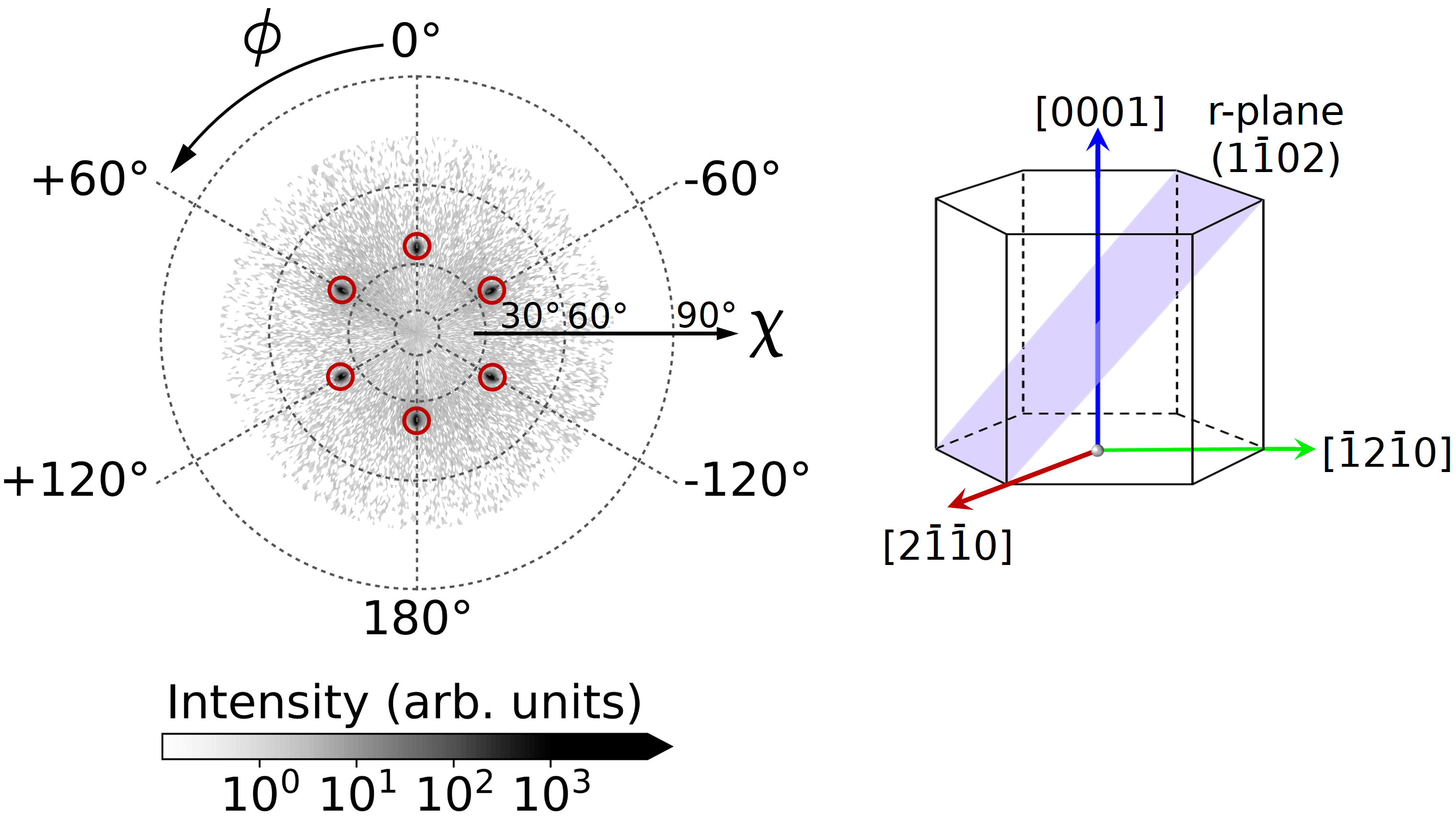}
    \caption{Pole figure for $\left\{1\bar102\right\}$ ($r$-planes) of PtSb(0001)/CrSb(0001) shown in stereographic projection. The sample was aligned such that \ce{SrF2}$[1\bar10]$ corresponds to $\phi=0\degree$. Six evenly spaced $1\bar102$ poles or diffraction maxima (circled in red) affirm the perfect 6-fold rotational symmetry of the (0001)-oriented hexagonal unit cells of both layers. The appearance of a pole at $\phi=0\degree$ implies PtSb/CrSb$[2\bar1\bar10]$ $\parallel$ \ce{SrF2}$[1\bar10]$. PtSb and CrSb cannot be distinguished in this measurement due to their nearly identical lattice constants. A sketch on the right illustrates the $r$-plane orientation within the unit cell.}
    \label{Pole figure}
\end{figure}

A pole figure measurement was made for the $\left\{1\bar102\right\}$ or $r$-planes of the PtSb/CrSb structure to determine its in-plane orientation and is illustrated in Fig.~\ref{Pole figure}. As expected from the lattice constants and hexagonal geometry, six $1\bar102$ poles are observed at $\rchi \approx38\degree$ (sample tilt) with a periodicity of 60$\degree$ in $\phi$ (sample azimuth). An epitaxial relationship of CrSb$[2\bar1\bar10](0001)$~$\parallel$~PtSb$[2\bar1\bar10](0001)$~$\parallel$~\ce{SrF2}$[1\bar10](111)$ is established based on the XRD results.

\begin{figure}
    \centering
    \includegraphics[width=0.7\textwidth, keepaspectratio]{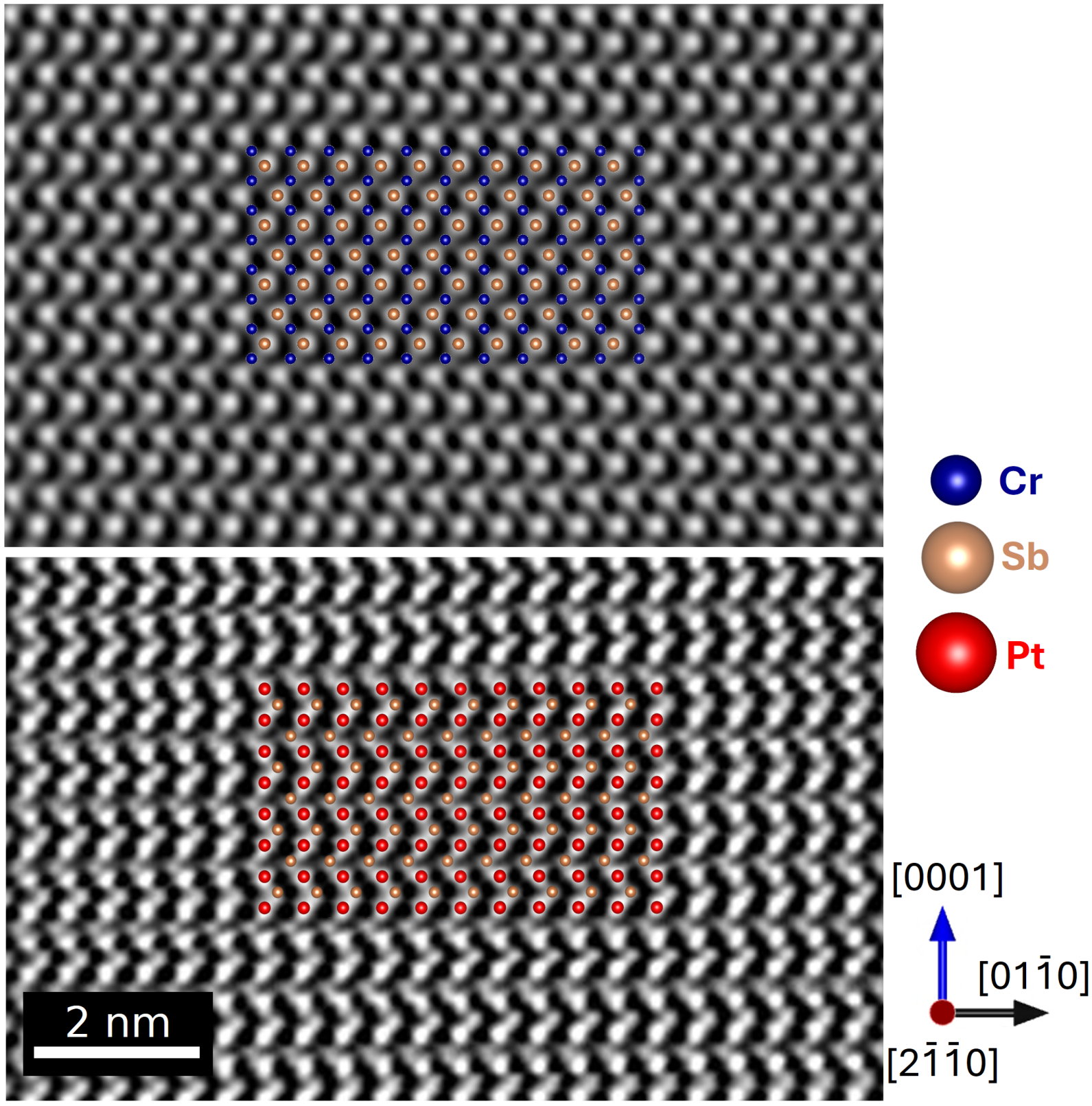}
    \caption{Cross-sectional HAADF-STEM of PtSb(0001)/CrSb(0001) showing their mutual epitaxial compatibility. The bilayer was grown on polished \ce{SrF2}(111). The corresponding crystal structure is overlaid on the images as a guide to the eye. Despite their isostructural nature, the two crystalline layers exhibit distinct Z-contrast due to the different atomic weights of Pt and Cr.}
    \label{TEM}
\end{figure}

HAADF-STEM of the cross-section of our heterostructure shows single crystalline layers with low inter-diffusion as suggested by the X-ray diffraction results. The atomic resolution imaging of the two single-crystalline films is shown in Fig.~\ref{TEM} and validates the continuation of the NiAs-type structure when Cr replaces Pt. Both layers are homogeneous over a large area with no observable crystallographic defects (see Fig.~S3 of Supplemental Material \cite{Supplemental_material} for \ce{SrF2}/PtSb and PtSb/CrSb interfaces).

\begin{figure}
    \centering
    \includegraphics[width=0.7\textwidth, keepaspectratio]{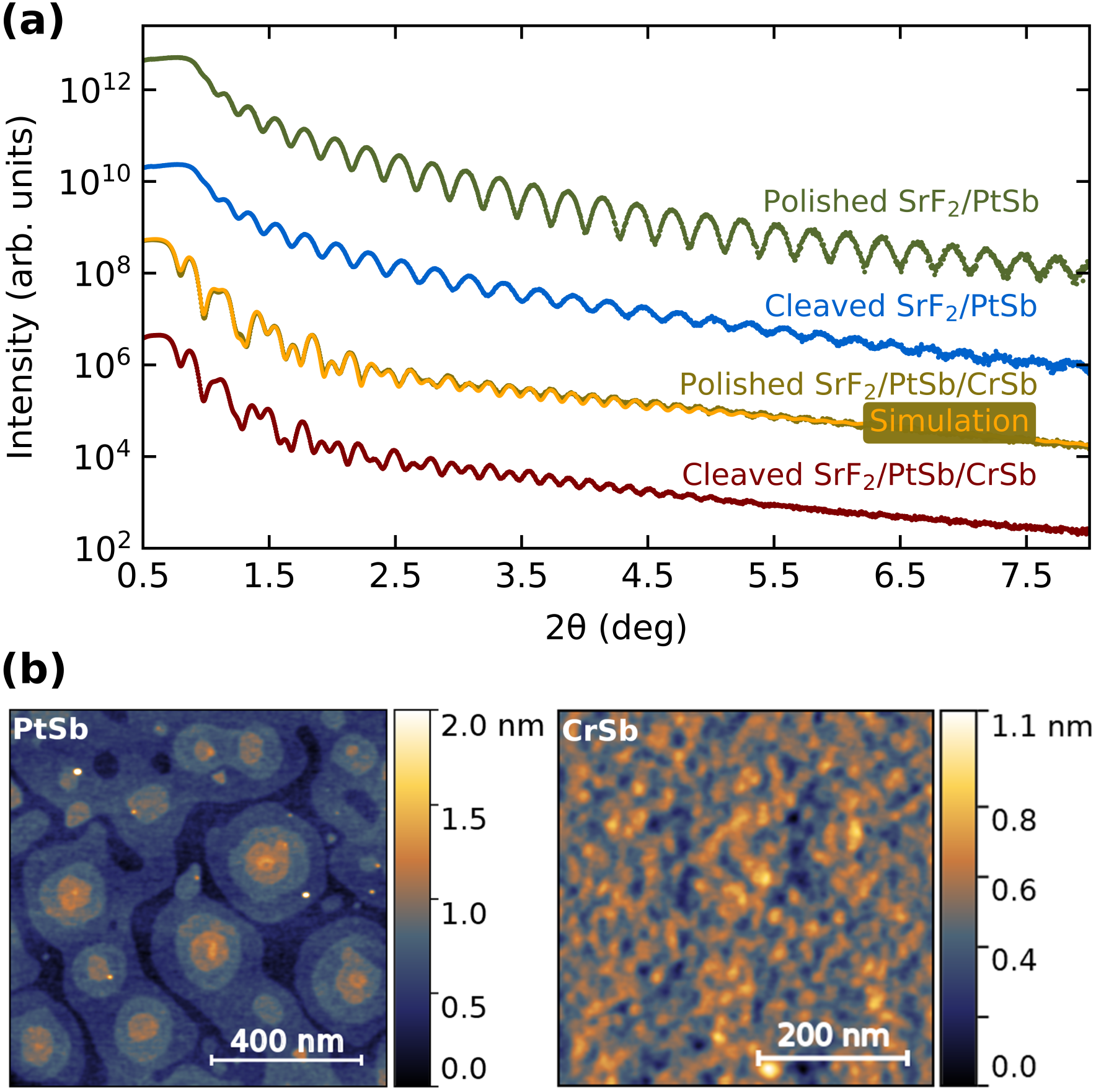}
    \caption{(a) XRR scans of \SI{30}{\nm} PtSb buffer and \SI{30}{\nm} PtSb/\SI{20}{\nm} CrSb bilayer on cleaved and polished \ce{SrF2}. A simulation for the heterostructure grown on a polished substrate is shown, with the simulation curve closely matching the experimental data, to the point of masking the experimental curve. The measurements were performed using Cu-K$\alpha_1$ radiation and are vertically offset for clarity. (b) Tapping mode AFM images showing the topography of \SI{30}{\nm} PtSb and \SI{30}{\nm} PtSb/\SI{20}{\nm} CrSb grown on cleaved \ce{SrF2}. Hexagonal faceted features can be seen on the PtSb(0001) surface.}
    \label{XRR}
    \label{AFM}
\end{figure}

While transmission electron microscopy investigates the surface and interface only on the nanoscale, X-ray reflectometry (XRR) allows us to conclude that the samples are also homogeneous on the macroscopic scale. XRR measurements of PtSb show pronounced Kiessig fringes that persist even at a large scattering angle of $10\degree$, indicating a very smooth film (Fig.~\hyperref[XRR]{\ref*{XRR}(a)}). The heterostructure data on polished substrates were matched by a model based on a \SI{30}{\nm} PtSb buffer and \SI{20}{\nm} CrSb layer using \emph{GenX 3} \cite{Glavic2022}. The layer densities obtained from the XRR fit correspond well with values reported for bulk crystals. Furthermore, one can determine a PtSb/CrSb interfacial roughness of $\approx$\SI{10}{\angstrom} and a surface roughness of $\approx$\SI{8}{\angstrom} (CrSb native oxide). Note that the sample on the cleaved substrate could not be fitted similarly due to the presence of large cleavage steps.

The low surface roughness of our films is also evident in atomic force microscopy (AFM) images. The surface of an optimal PtSb film (Fig.~\hyperref[AFM]{\ref*{AFM}(b)}) reveals atomically flat terraces of \SI{\approx100}{\nano\meter} width, establishing it as an ideal template for heteroepitaxial growth. These terraces are separated by biatomic steps, causing the film to terminate in the same atomic plane across the whole surface. Though similar topographic features are absent in the CrSb film, it is smooth with a root mean square (RMS) roughness $<$~\SI{2}{\angstrom}. Note that this value differs from the roughness observed in XRR measurements mostly due to the fact that a cleaved substrate was used for the samples studied by AFM imaging.

\begin{figure}
    \centering
    \includegraphics[width=0.6\textwidth, keepaspectratio]{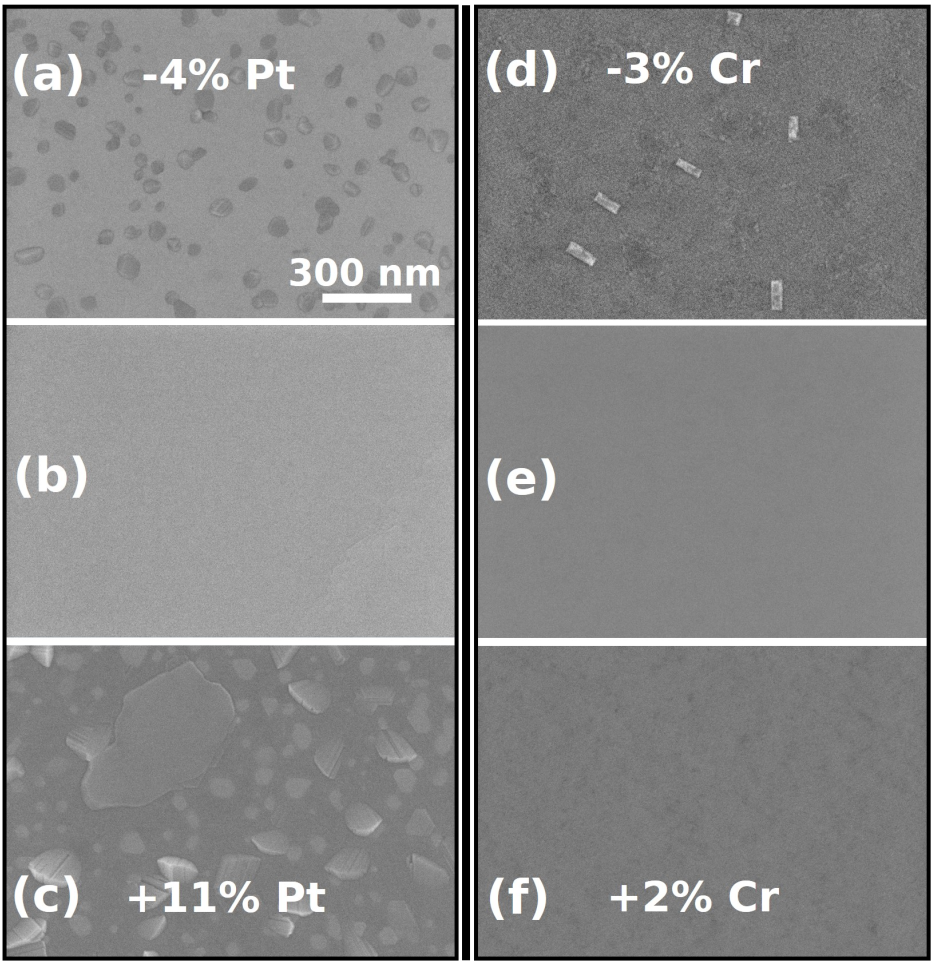}
    \caption{SEM images showing the dependence of thin film morphology on composition. The images were captured using an in-lens detector. PtSb (a to c): 4 at.\% Pt deficit, close to ideal, 11 at.\% Pt excess, respectively. CrSb (d to f): 3 at.\% Cr deficit, close to ideal, 2 at.\% Cr excess (seen as dark patches with a weak contrast), respectively.}
    \label{SEM}
\end{figure}

Scanning electron microscopy (SEM) was used to optimize the elemental composition of both layers while also checking their topography. Optimized films are observed to be smooth and featureless while even a small excess of Pt, Cr or Sb results in precipitates as seen in Fig.~\ref{SEM}. The chemical composition of our thin films was investigated using an electron probe microanalyzer (EPMA) equipped with a wavelength-dispersive X-ray spectrometer. The stoichiometry was checked at multiple points on each sample with a probing area of $\approx{\SI{1}{\micro\meter}^2}$ and was confirmed to be homogeneous across the film. Compositions of $48.4~\pm~1\,\text{at.\%}$ Pt and $51.6~\pm~1\,\text{at.\%}$ Sb, as well as $50.9~\pm~1\,\text{at.\%}$ Cr and 49.1$~\pm~1\,\text{at.\%}$ Sb were measured for optimized PtSb and CrSb, respectively. The stoichiometry of CrSb was obtained from a layer grown on a bare sapphire substrate at conditions otherwise equal to the deposition parameters of the heterostructure. These results were supported by wavelength-dispersive X-ray fluorescence (XRF) measurements, which yielded similar values of $48.7~\pm~3\,\text{at.\%}$ Pt, $51.3~\pm~3 \,\text{at.\%}$ Sb and 48.7$~\pm~3\,\text{at.\%}$ Cr, 51.3$~\pm~3\,\text{at.\%}$ Sb. Overall, the chemical analysis suggests that our optimized samples showing a smooth surface have close to ideal/stoichiometric compositions.

\begin{figure}
    \centering
    \includegraphics[width=0.8\textwidth, keepaspectratio]{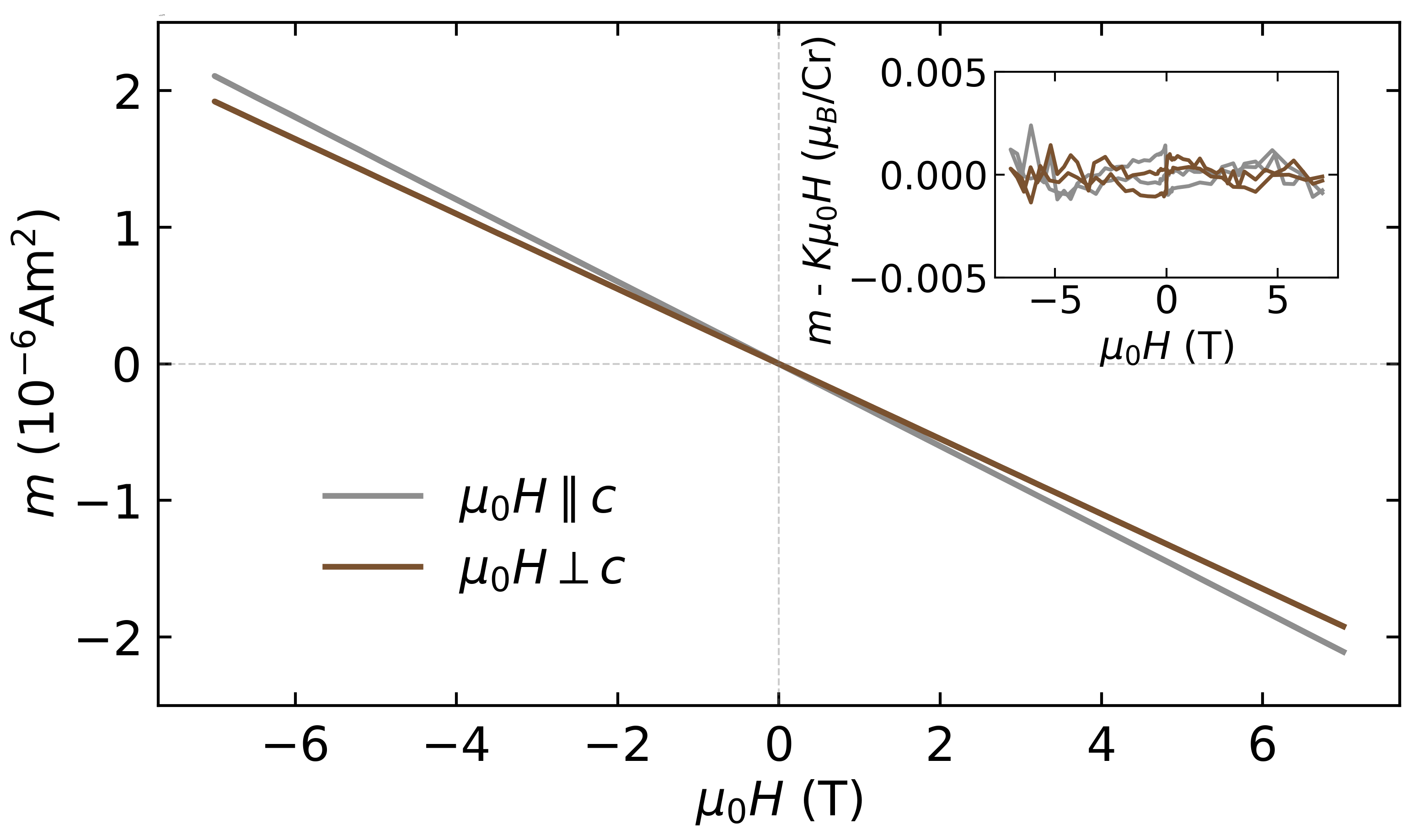}
    \caption{Magnetic moment $m$ vs. field $H$ of a \ce{SrF2}/\SI{30}{\nm} PtSb/\SI{20}{\nm} CrSb sample using SQUID magnetometry. The signal is dominated by the diamagnetic substrate. The inset displays the same data with an line of slope $K$ subtracted, leaving only noise behind.}
    \label{SQUID}
\end{figure}

Superconducting quantum interference device (SQUID) magnetometry was used to study the magnetic properties of our optimized heterostructure. The measured magnetic moment is dominated by substrate diamagnetism as shown in Fig.~\ref{SQUID}. The observed response can be interpreted as a sum of the diamagnetic contribution from \ce{SrF2}, an unknown contribution (diamagnetic or paramagnetic) from the PtSb buffer and the positive susceptibility of altermagnetic CrSb. A vanishingly small magnetic moment is found upon subtracting a linear function from this data, consistent with the behavior expected from a compensated magnet. Within the limitations of our method and the associated uncertainty, we assign an upper limit for the moment in zero magnetic field to $\SI{0.001}{\mu}_B$/Cr.

\begin{figure}
    \centering
    \includegraphics[width=0.7\textwidth, keepaspectratio]{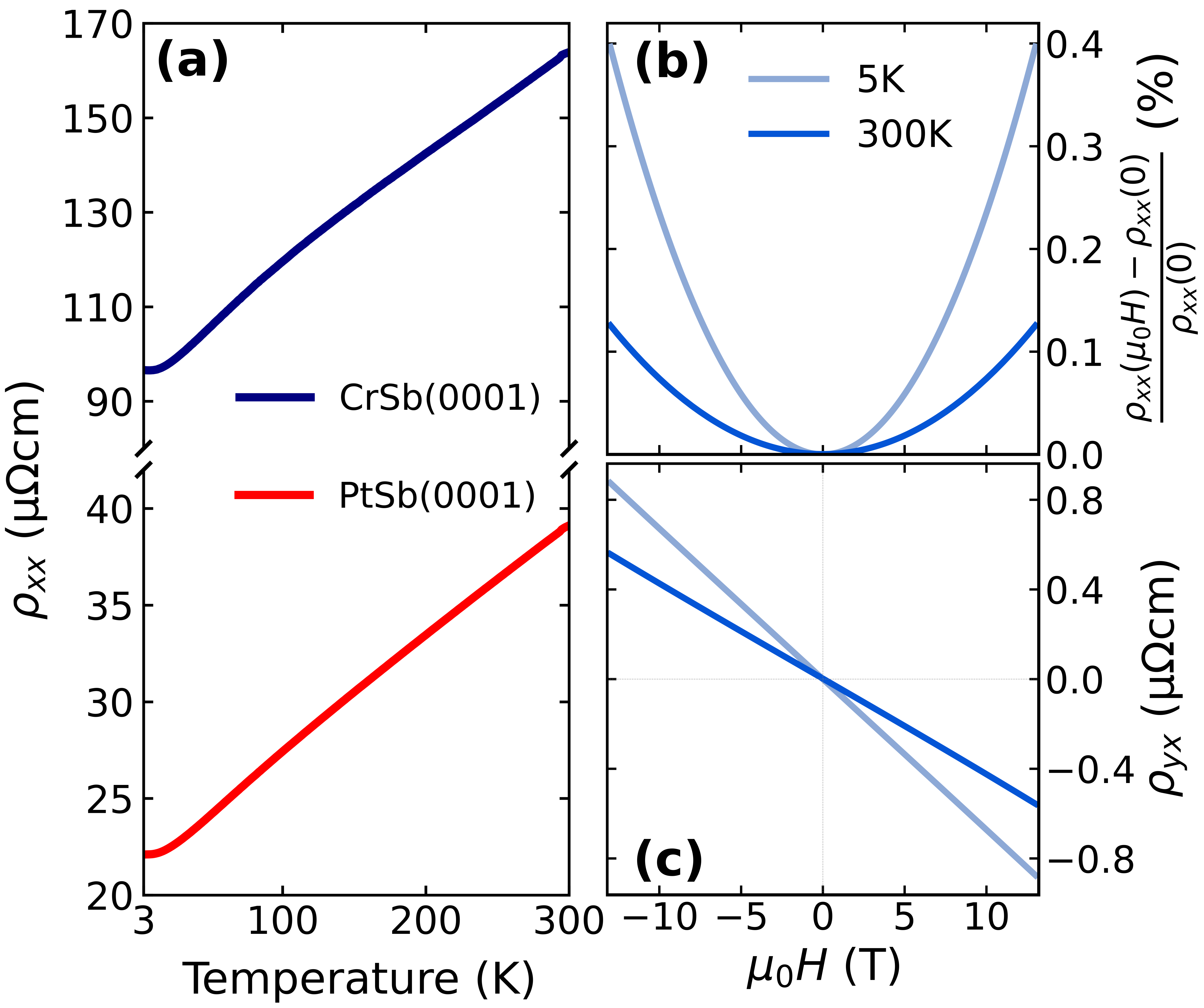}
    \caption{(a) Longitudinal resistivity $\rho_{xx}$ vs. temperature of epilayers of PtSb and CrSb. (b,c) Magnetoresistance and Hall resistivity $\rho_{yx}$ of CrSb  for magnetic field applied along the [0001] direction at 5 and \SI{300}{\kelvin}. The data for CrSb was extracted from that of a \SI{15}{\nm} PtSb/\SI{25}{\nm} CrSb bilayer assuming a parallel resistors circuit model.}
    \label{Rho}
\end{figure}

The magnetotransport properties of these thin films were studied in Hall bars, which were defined by optical lithography combined with Ar$^{+}$ ion milling.
Though the growth parameters were initially optimized using a
\SI{30}{\nm} PtSb buffer, the PtSb thickness was reduced in the transport samples to enhance the contribution of the CrSb layer to the data. We found that the single crystallinity in our heterostructure is retained even upon decreasing the buffer thickness down to \SI{10}{\nano\meter} (Supplemental Material Fig. S4 \cite{Supplemental_material}). For the transport results shown here a \SI{15}{\nm} PtSb/\SI{25}{\nm} CrSb bilayer, and a \SI{15}{\nm} PtSb single layer were investigated to allow for separation of the PtSb and CrSb transport contributions. The resistance of these structures was measured as a function of temperature using the four-point probe method and the separation into PtSb and CrSb signals was performed assuming a parallel resistor model. The resulting longitudinal resistivity, as presented in Fig.~\hyperref[Rho]{\ref*{Rho}(a)}, shows typical metallic behavior for both PtSb and CrSb with residual resistance ratios (RRR) of $\approx1.8$ and 1.7, respectively. Magnetoresistance and Hall resistance measurements for magnetic field applied perpendicular to the film plane are analyzed using the parallel resistor model while also considering the shunting of transversal voltage in the bilayer (see Supplemental Material \cite{Supplemental_material} and \cite{Xu2008, Arnaudov2003} for details). The results extracted for CrSb show small positive magnetoresistance commonly found in metals and a strictly linear Hall resistivity consistent with the expected absence of the anomalous Hall effect as discussed further below. When analyzed using a single-band model one finds hole carrier densities of \SI{9e21}{\per\cubic\centi\meter} with a Hall mobility of \SI{7}{\centi\meter\squared\per\volt\per\second} at \SI{5}{\kelvin}. The predominant hole-like transport was confirmed from a second set of bilayer and reference samples with a different CrSb:PtSb thickness ratio. Our results are consistent with reports on bulk CrSb \cite{Urata2024, Bai2025}.

\section{Discussion}

Using magnetron sputtering, we have achieved CrSb films on a \ce{SrF2}/PtSb template that are both smooth and epitaxial with unprecedented crystal quality. 
This was achieved by the comparatively inert \ce{SrF2} substrate, which simplifies surface preparation and inhibits chemical reactions with the deposited material. This is in contrast to growth on \mbox{III-Vs}, despite their promising lattice match, as we detail below.

\mbox{III-V} (111) surfaces of InP and GaAs have a lattice mismatch of only $\approx0.7\%$ and $\approx3\%$ with respect to CrSb. The isostructural MnTe, with a similar in-plane lattice spacing ($a =$ \SI{4.14}{\angstrom}), was successfully grown on InP(111)A by MBE \cite{Kriegner2016, Amin2024}. However, our attempts of CrSb deposition on InP and GaAs led to polycrystalline films regardless of substrate termination. The (0001) orientation of CrSb can be promoted by an increased growth temperature on InP(111)A, but this induces an exchange reaction of Sb with P, forming InSb. Sb generally replaces P in InP and As in GaAs (all group V elements), forming In and Ga antimonides, consistent with previous observations \cite{Nakai2005, MOUSLEY2018, Jablonska2009, HATFIELD2006}. This also limits the deposition temperature of antimonides on GaAs, where we find GaSb formation in our growth attempts of PtSb on the GaAs(111)B surface. The problem of chemical reactions with \mbox{III-V} substrates is absent in the case of tellurides as Te is a group VI element.

He et al. reported epitaxial CrSb on GaAs(111)B via MBE \cite{He2017, He2018}, with 3D island morphology and low crystal quality. In contrast, our sputtered films on GaAs(111)B are continuous and smooth, but remain polycrystalline. We note that Reimers et al., who also used dc magnetron sputtering (from a multi-segmented Cr/Sb target), obtained single-crystalline CrSb with mosaicity of $0.5\degree$, but in the ($10\bar10$) ($m$-plane) orientation by switching to GaAs(110) \cite{Reimers2024}.

\ce{SrF2}(111) was successfully used for epitaxial growth of MnTe(0001) \cite{Kriegner2017}. \ce{SrF2} neatly circumvents issues associated with native oxide removal and Sb reactivity. However, our direct growth of CrSb on \ce{SrF2} resulted in polycrystalline films with more crystallite orientations than those observed on \mbox{III-Vs}, however, with absence of the desired (0001) orientation.

Buffer layers can serve as an alternative strategy to achieve single-crystalline growth with the desired crystalline orientation. For CrSb, Burrows et al. previously successfully grew the (0001) orientation using a \SI{100}{\nm} buffer of iso-structural MnSb(0001) on GaAs(111) \cite{BURROWS2019}. However, no information on their crystalline quality or phase purity was provided. Unfortunately, such a ferromagnetic underlayer makes it extremely challenging to disentangle the altermagnetic properties of CrSb, making this approach less favorable. Similarly, He et al. showed that the morphology of CrSb on GaAs(111)B could be improved by inserting a \SI{2}{\nano\meter} buffer layer of the topological insulator \ce{(Bi,Sb)2Te3} \cite{He2017, He2018}.

Instead, we explored PtSb as a buffer since it is isostructural and found to nucleate as a single-crystal with (0001) orientation on \ce{SrF2}. This non-magnetic buffer allows for an easier study of CrSb's compensated magnetic order and magnetotransport properties. The structural quality of these films is crucial for investigating their intrinsic behavior, particularly in the context of neutron diffraction and AHE.

Regarding the AHE, we note that the altermagnetic $g$-wave nature of collinearly ordered NiAs-structured materials permits an AHE depending on the magnetic moment orientation \cite{Betancourt2023, Betancourt2024}. In MnTe, moments aligned along $[1\bar100]$ support a finite AHE, whereas the $[0001]$ moment direction connected with the $6'/m'mm'$ magnetic point group symmetry in hexagonal CrSb does not. This aligns with the linear Hall response observed in our work, which we attribute to the ordinary Hall effect. In contrast, bulk CrSb crystals measured in the same geometry exhibit strong nonlinearities, which could still be explained by multi-band transport within the ordinary Hall effect framework \cite{Urata2024, Bai2025}. Since laboratory fields cannot reorient the moments, Zhou et al. induced symmetry breaking via strain, leading to a finite altermagnetic AHE \cite{Zhou2025}.

\section{Conclusion}
Epitaxial CrSb(0001) thin films of a very high quality have been successfully fabricated by dc magnetron co-sputtering. This was made possible by using a PtSb(0001) underlayer grown on the \ce{SrF2}(111) surface. The tendency of PtSb to grow in a single-crystalline manner ensured the same for the iso-structural CrSb overlayer. The films were also observed to be smooth and show typical metallic conductivity. The high crystalline quality achieved by our method could be instrumental in the study of spin-split altermagnetic bands via angle-resolved photoemission spectroscopy (ARPES) and the investigation of magnetic structure in thin CrSb films using neutron diffraction. PtSb could also be used to seed the high-quality epitaxial growth of other materials interesting for spintronics, such as MnSb and MnTe. NiAs-type crystals exhibit a wide range of magnetic structures as well as transport properties and often show mutual epitaxial compatibility. Thus, altermagnetic CrSb and MnTe could be combined with other interesting compounds from this class to explore the rich physics in such systems.

\section*{Acknowledgements}

We acknowledge financial support by the Czech Science Foundation (Grant No. 22-22000M), Lumina Quaeruntur fellowship LQ100102201 of the Czech Academy of Sciences, Czech Ministry of Education, Youth and Sports (MEYS) grants LM2023051 and CZ.02.01.01/00/22\_008/0004594, ERC Advanced Grant No. 101095925 and the Dioscuri Program LV23025 funded by Max Planck Society and MEYS. We are grateful to Pavel~Machek, Zbyněk~Šobáň, Jiří~Zyka, Vlastimil~Jurka, Karel~Hruška and Veronika~Růžičková from the Institute of Physics, Czech Academy of Sciences and ScopeM, ETH Zurich for technical support and assistance.

%\bibliography{ref}

\end{document}